\documentclass[twocolumn,preprintnumbers,amsmath,amssymb]{revtex4}

\usepackage{graphicx}
\usepackage{dcolumn}
\usepackage{bm}
\usepackage{wasysym}
\usepackage{subfigure}
\usepackage{float}

\newcolumntype{C}[1]{>{\centering\let\newline\\\arraybackslash\hspace{0pt}}m{#1}} 

\begin{document}

\preprint{}

\title{Electronic and Structural Properties of Vacancies and Hydrogen Adsorbates on Trilayer Graphene}

\author{Marcos G. Menezes$^1$}
 \email{marcosgm@if.ufrj.br}
\author{Rodrigo B. Capaz$^{1}$}%
\affiliation{$^1$ Instituto de F\'{i}sica, Universidade Federal do Rio de Janeiro, Caixa Postal 68528 21941-972, Rio de Janeiro, RJ, Brazil}

\date{\today}

\begin{abstract}
\b
Using \textit{ab initio} calculations, we study the electronic and structural properties of vacancies and hydrogen adsorbates on trilayer graphene. Those defects are found to share similar low-energy electronic features, since they both remove a $p_z$ electron from the honeycomb lattice and induce a defect level near the Fermi energy. However, a vacancy also leaves unpaired $\sigma$ electrons on the lattice, which lead to important structural differences and also contribute to magnetism. We explore both ABA and ABC stackings and compare properties such as formation energies, magnetic moments, spin density and the local density of states (LDOS) of the defect levels. These properties show a strong sensitivity to the layer in which the defect is placed and smaller sensitivities to sublattice placing and stacking type. Finally, for the ABC trilayer, we also study how these states behave in the presence of an external field, which opens a tunable gap in the band structure of the non-defective system. The $p_z$ defect states show a strong hybridization with band states as the field increases, with reduction and eventually loss of magnetization, and a non-magnetic, midgap-like state is found when the defect is at the middle layer.
\end{abstract}

\maketitle

\section{Introduction\label{sec:introduction}}

Graphene, a 2D material composed of carbon atoms arranged in a honeycomb lattice, has attracted much attention from the scientific community in the last decade due to its unique electronic and structural properties and its potential applications in a wide variety of fields \cite{neto_rmp, novoselov_nature}. Among many topics studied in graphene research, one of particular interest is the study of defects, and how they affect the electronic and structural properties of the sheet. In particular, two well-studied types of defects in graphene are vacancies and hydrogen adsorbates on top of the sheet, which can be produced experimentally by bombarding the sheet with high-energy particles such as protons \cite{esquinazi_prl, ruffieux_prl}. Such defects generally induce a local magnetism in the graphene sheet, a highly desirable property for spintronics applications \cite{oleg_prb, oleg_rpp, nair_natcom}. Magnetic materials commonly used in current technological applications are based on heavy elements, with $d$ and $f$ orbitals, so the realization of $s$ and $p$ electron magnetism in graphene is very attractive.

In single layer graphene, isolated vacancies and hydrogen adsorption are found to have similar low-energy properties. Both defects remove a $p_z$ electron from the honeycomb lattice, inducing a "quasi-localized" defect level near the Fermi energy. The associated wavefunction is found to be long-ranged, decaying as $1/r$, where $r$ is the distance to the defect center, as predicted by theory \cite{pereira_prl, nanda_njp} and revealed by STM experiments on vacancies in graphite \cite{ugeda_prl}.

\begin{figure}[h]
\centering
\includegraphics[width=7.0cm]{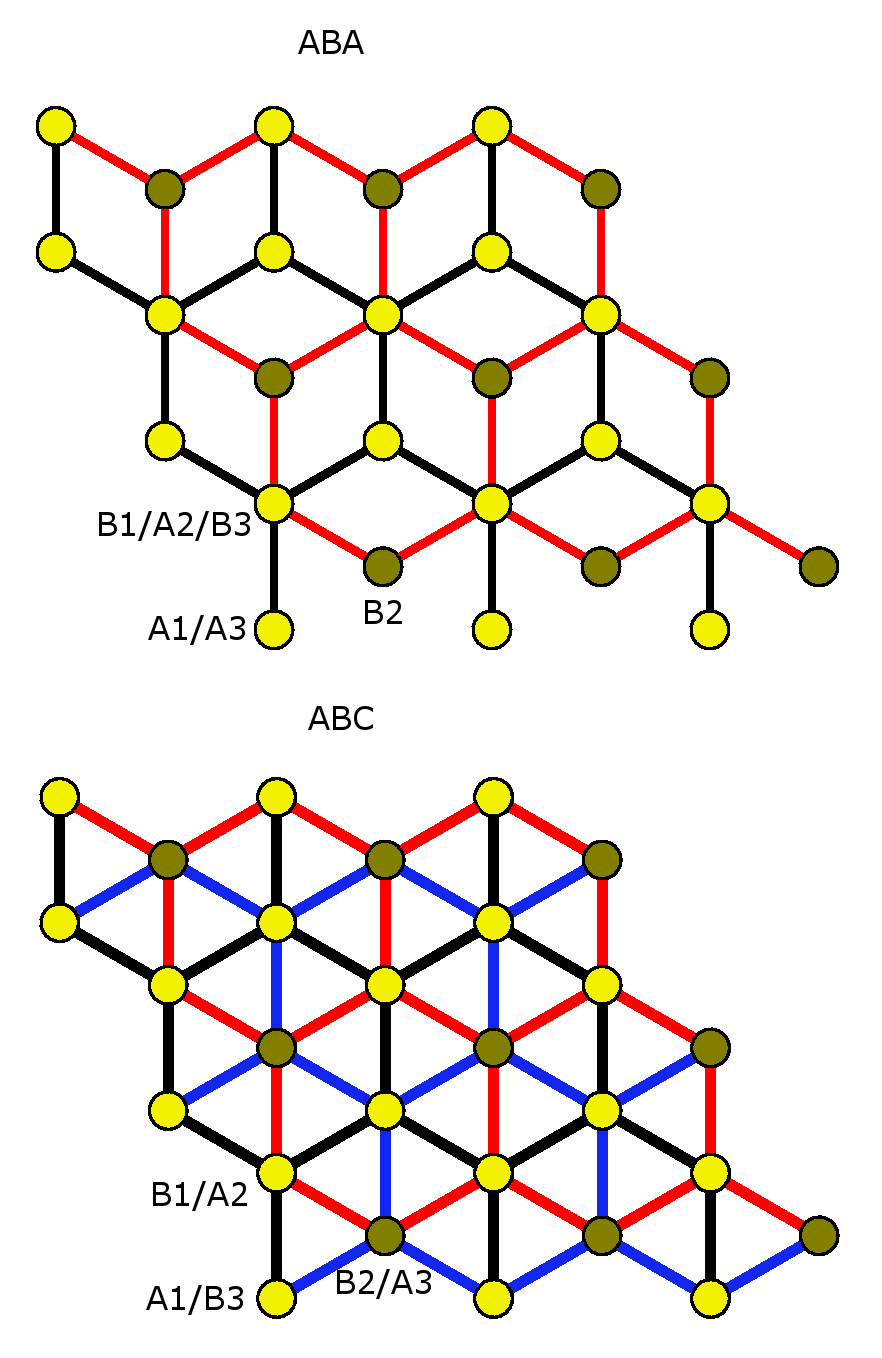}
\caption{\label{fig:trilayers} Top view of ABA (top) and ABC (bottom) trilayer graphene. Yellow and dark yellow spheres represent carbon atoms on top and middle layers, respectively. The basis atoms are labeled by their sublattice (A or B) and layer indexes (1, 2 or 3). Sites with more than one index indicate different carbon atoms that are on top of each other. The bonds in the top, middle and bottom layers are represented by the black, red and blue lines, respectively (blue is only visible in the ABC trilayer).}
\end{figure}

Pure trilayer graphene also attracts increasing attention due to its stacking-dependent electronic properties. The two main types of stacking, ABA and ABC, are shown in Fig. \ref{fig:trilayers}. In ABA (or Bernal) stacking, the top and bottom layers have the same orientation, while the middle layer is rotated by $60^{\circ}$ with respect to them. The associated low-energy band structure shows a pair of single-layer and bilayer-like bands, with linear and quadratic dispersions, respectively \cite{expABA, expABC, McCannABA, menezes_prb}. On the other hand, in ABC stacking, all the layers are rotated by the same angle with respect to each other, so they all have different orientations. The corresponding low-energy band structure in this case shows a pair of bands with cubic dispersion. In addition, the application of an external electrical field perpendicular to the sheets in ABC trilayer breaks its inversion symmetry and induces a tunable band gap, a very desirable property for electronic devices \cite{expABC, HallABC, MacDonaldABC}. Point defects in trilayer graphene are expected to give rise to a wider variety of electronic and structural features, since they might depend on the stacking order and the site where the defect is.

In this work, we explore these different possibilities, by employing \textit{ab-initio} calculations based on Density Functional Theory (DFT). We compare the features of the low-energy defect level induced by an isolated vacancy and by adsorption of a single hydrogen atom on top of ABA and ABC trilayer graphene. Specifically, we place the defect at different non-equivalent sites and study how properties such as formation energies, magnetizations, band structure, spin densities and local density of states (LDOS) behave in each case. For the ABC trilayer, we also study the effect of an external electrical field, where we look in particular for the possibility of a mid-gap defect level, as already pointed out by a previous tight-binding calculation \cite{castro_sscomm}. Our paper is organized as follows: in the next Section, we present the methodology used in this work. In Sections \ref{sec:hydrogen} and \ref{sec:vacancy}, we discuss our results for hydrogen adsorption and isolated vacancies, respectively, where the properties mentioned above are discussed in different subsections. Finally, in Section \ref{sec:conclusions}, we summarize our results and present our conclusions.

\section{Methodology\label{sec:methodology}}

We use first-principles calculations based on the Density Functional Theory (DFT) \cite{ho-kohn, kohn-sham}, as implemented in the SIESTA code \cite{siesta}. We use norm-conserving Troullier-Martins pseudopotentials \cite{troullier-martins} for the ion-electron interaction, a PBE-GGA exchange-correlation functional \cite{perdew-burke-ernzerhof} and a double-$\zeta$-polarized (DZP) pseudoatomic basis set for the expansion of the electronic wavefunctions. The real space grid energy cutoff is set to $150$ Ry and we use a $6 \times 6 \times 1$ Monkhorst-Pack k-point grid \cite{monkhorst-pack} and an electronic smearing (Fermi-Dirac-like function) of $100$ K in all calculations.

For all configurations considered, we let the atomic coordinates relax in all directions until all the forces are smaller than $0.04$ eV/\AA. However, for the cases where an electrical field is applied (ABC trilayer only), we use the optimized coordinates from the corresponding configuration in zero field. In order to simulate the vacancy and the hydrogen adsorption defects, we use supercells of dimension $6 a \times 6 a$, where $a = 2.47$ \AA \ is the lattice constant of pure graphene. This corresponds to a defect density of $\sim 5 \times 10^{13}$ cm$^{-2}$, which is a high density compared to current experimental values \cite{chen_prl}. The optimized value for the interlayer distance is $3.27$ \AA \ in all cases and we use a vacuum of $10$ \AA \ in the direction perpendicular to the layers in order to avoid interactions between periodic images \footnote{Even though we are using a GGA-PBE functional in our calculations, the optimized value for the interlayer is smaller than the experimental value. We believe this result might be related to the localized character of the DZP basis used to expand the wavefunctions in SIESTA, which may contribute to overbinding. Nevertheless, we find that small variations in the interlayer distance do not play any major role for most defect configurations, as we discuss later.}. We have tested supercells of larger sizes for a few configurations and we didn't see qualitative changes of our results. The different configurations studied in this work are labeled by the stacking type (ABA or ABC), the type of defect and the site where the defect is, following the notation of Fig. \ref{fig:trilayers}. For hydrogen adsorption, the H atom always lies on top of the nearest carbon atom and the bond length is relaxed together with the the structure. For A1 and B1 sites, we always choose the hydrogen atom to be outside, instead of between the layers as in the A2 and B2 sites. We performed both spin polarized and unpolarized calculations, but we report mainly our results for the spin polarized case, since they are of more interest, unless mentioned otherwise.

\section{Hydrogen Adsorption\label{sec:hydrogen}}

\subsection{Structural properties, formation energies and magnetic moments \label{sec:h:struct}}

Our calculations show that the structural changes due to hydrogen adsorption are small and nearly stacking-independent. The hydrogen atom "pulls" the carbon atom it binds to from the layer, along with some of its neighbors. This effect is due to a local $sp^2 - sp^3$ rehybridization in order to form the C-H bond. As shown in Table \ref{tab:H}, the relaxed length of the C-H bond lies between $1.13 - 1.15$ \AA \ in all cases. However, the magnitude of the displacement of the carbon atoms depends on the defect site. When the H atom is on top of A1 or B1 sites (outside the trilayer), the out-of-plane displacement of the C atom in the C-H bond ranges from $0.41$ to $0.49$ \AA, while its nearest neighbors move between $0.07 - 0.14$ \AA. On the other hand, when the H atom is on top of A2 and B2 sites, between the top and middle layers, the displacements are smaller: between $0.20 - 0.28$ \AA \ and $-0.02 - 0.00$ \AA, respectively. We believe this effect is related to the interlayer interaction: in the first case, the carbon atoms have more freedom to relax, while in the second case they are tightly squeezed between the two adjacent layers, leading to a smaller displacement. We also do not see any changes in the interlayer distances due to the presence of the hydrogen atom. Finally, we do not see any noticeable structural changes in the adjacent pristine layers.

\begin{table*}
\caption{\label{tab:H} Results for spin-polarized calculations of hydrogen adsorption on trilayer graphene: C-H bond length ($d_{C-H}$), out-of-plane displacements of the carbon atom in the C-H bond ($d_{C_H}$) and its first neighbors ($d_{C_1}$), formation energies ($E_f$) and magnetic moments per unit supercell ($M$) for each configuration studied. For $d_{C_1}$, negative values represent displacements in a direction opposite to the H atom.}
\begin{tabular*}{\textwidth}{  *{6}{C{2.6cm}} }
\hline
Defect Site & $d_{C-H}$ (\AA) & $d_{C_H}$ (\AA) & $d_{C_1}$ (\AA) & $E_f$ (eV) & $M$ ($\mu_B$) \\
\hline
\multicolumn{6}{c}{ABA trilayer} \\
\hline
A1 & 1.13 & 0.41 &  0.07 & 1.14 & 1.00 \\
A2 & 1.15 & 0.20 & -0.01 & 1.37 & 0.00 \\
B1 & 1.13 & 0.45 &  0.11 & 1.16 & 1.00 \\
B2 & 1.13 & 0.25 & -0.01 & 1.23 & 0.01 \\
\hline
\multicolumn{6}{c}{ABC trilayer} \\
\hline
A1 & 1.13 & 0.48 &  0.13 & 1.11 & 0.91 \\
A2 & 1.14 & 0.24 & -0.02 & 1.31 & 0.51 \\
B1 & 1.14 & 0.49 &  0.14 & 1.17 & 1.04 \\
B2 & 1.13 & 0.28 &  0.00 & 1.29 & 0.87 \\
\hline
\end{tabular*}
\end{table*}

In order to further study the sensitivity to the defect sites, we calculate the formation energies and magnetic moments for each configuration. For hydrogen adsorption, the formation energy is calculated as
\begin{equation}\label{eq:form_en_H}
E_f = E_{trilayer + H} - E_{trilayer} - \frac{E_{H_2}}{2},
\end{equation}
where $E_{trilayer + H}$ is the total energy of a given configuration, $E_{trilayer}$ is the total energy of the corresponding pristine trilayer (with the same supercell size) and $E_{H_2}$ is the total energy of an isolated hydrogen molecule, calculated using a large cubic supercell. The results are also reported in Table \ref{tab:H}. As expected, the formation energies for H adsorption in middle layer sites are larger than those for the external layers. The highest value is found for the A2 site in the ABA trilayer, since this site has two carbon atoms on top of it, one in each external layer. In the ABC trilayer, the A2 and B2 sites have similar environments, and the energies differ slightly by the placement of the H atom: between two C atoms in the former and between a C atom and a hexagon center in the latter. Finally, for both trilayers, the formation energies for the B1 site are larger than those of the A1 site, again due to the presence of a C atom on the top of the B1 site. For calculations without spin polarization, the formation energies are systematically larger by a few tenths of meV, and follow the same trends of Table \ref{tab:H}. The structural parameters, formation energies and magnetizations reported in Table \ref{tab:H} are qualitatively consistent with previous calculations on single and bilayer graphene \cite{boukhvalov_prb, casolo_jchem, arellano_jchem, oleg_prb}. On the other hand, a sensitivity of these properties to the defect placement is not seen in single layer graphene (due to the equivalence of the A and B sublattices) and there is no layer sensitivity in the bilayer (due to its inversion symmetry).

The magnetization shows the most striking differences between the trilayers. For the ABA trilayer, when the defect is on one of the middle layer sites, there is no net magnetization in the supercell after the structural relaxation. On the other hand, for the external layer sites, a "full" magnetization is observed, coming from a full occupation of one of the spin-split defect levels. For the ABC trilayers, the A1, A2 and B2 sites show partial magnetization, while the B2 site retains full magnetization. We now discuss these results in terms of the band structure.

\subsection{Band structure \label{sec:h:bands}}

\begin{figure*}
\centering
\includegraphics[width=\textwidth]{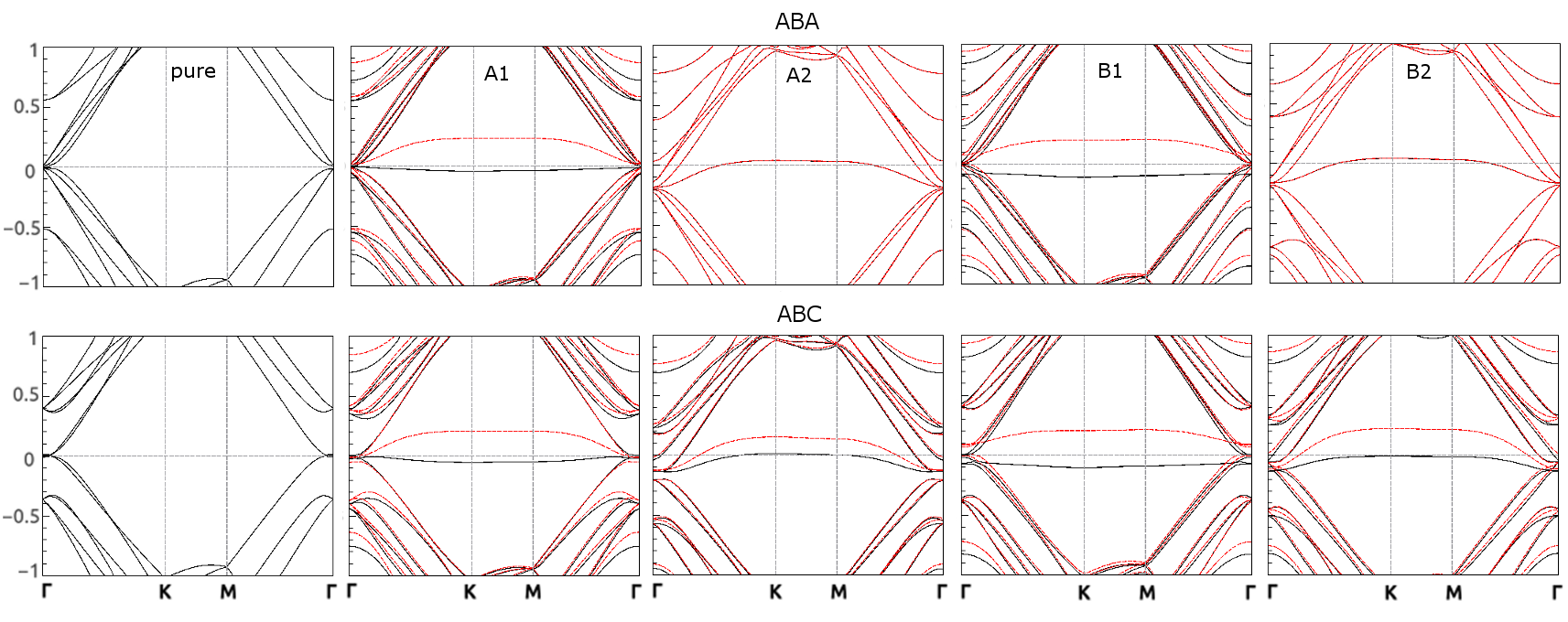}
\caption{\label{fig:h:bands} Band structure results for hydrogen adsorption on ABA (top) and ABC (bottom) trilayer graphene. The H atom is on top of the site indicated in each frame. Spin up (down) levels are represented by black (red dashed) lines and the Fermi level is set to zero in all cases. Energies are in eV. For the pure trilayers (first column), the calculation is performed without spin polarization, so only black lines are shown.}
\end{figure*}

The band structures for each defect configuration are shown in Fig. \ref{fig:h:bands}. We also included the band structure for the pristine ABA and ABC trilayers, labeled "pure" in the corresponding frames. In a $6 \times 6$ supercell, the $K$ and $K'$ points from the primitive Brillouin Zone (BZ) of pure trilayer graphene are both folded into the $\Gamma$ point of the supercell BZ, hence many double degeneracies are observed in the bandstructure in these cases. The introduction of a defect in the supercell lifts most of these degeneracies and also introduces a new defect level near the Fermi energy, which has a small bandwidth and is mostly flat throughout the BZ. The high density of states associated with such a level leads to a strong exchange interaction and a spin-split occurs, the so called Stoner effect, as is observed in most cases. We now discuss each case in more detail.

In the ABA trilayer, the band structure near the Fermi energy is composed of a pair of bands with linear dispersion and two pairs of bands with quadratic dispersion, similar to those observed in single layer and bilayer graphene, respectively. Due to the lack of inversion symmetry in this trilayer (it only has mirror symmetry), both type of bands have small gaps of a few tenths of meV. With the adsorption, we observe that the gap between the linear bands is increased to about $0.1$ eV in all cases and the linear behavior is lost in a greater vicinity of the $\Gamma$ point. On the other hand, the gap between the closest quadratic bands is preserved.

\begin{figure}[H]
\centering
\includegraphics[width=8.5cm]{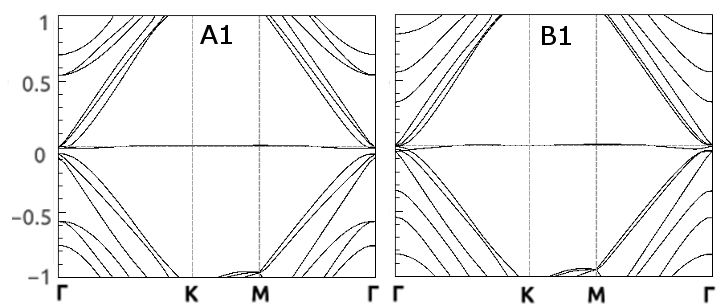}
\caption{\label{fig:h:bands:nm} Band structures without spin polarization for hydrogen adsorption on top of A1 and B1 sites in the ABA trilayer. The defect levels are much flatter in these cases, thus favoring spin polarization, as can be seen in Fig. \ref{fig:h:bands}.}
\end{figure}

In the A2 and B2 cases, where the defect is bound to the middle sheet, the band structure does not show spin polarization, while in the A1 and B1 cases a spin-split occurs, which is specially large in the B1 case. This behavior can be understood by looking at the bandwidth of the defect level in a corresponding non-polarized calculation. For the A1 and B1 sites (Fig. \ref{fig:h:bands:nm}), our calculations show that the induced defect bands are much flatter than those induced in the A2 and B2 cases shown in Fig. \ref{fig:h:bands}. Therefore, the density of states at the Fermi energy in these cases is larger and the Stoner criterion for exchange splitting is satisfied. However, we point out that the bandwidth of the defect level in the A2 and B2 cases is very sensitive to the interlayer distance, since the H atom lies between two adjacent sheets. To check this effect, we performed test calculations with variations of $\pm 5\%$ in this distance and we found that a spin-split can also be observed in these cases, in a similar fashion to the ABC trilayer, with fractional magnetizations of up to $0.8 \mu_B$  \footnote{Another way in which a magnetic state could also be achieved in the A2 and B2 configurations of the ABA trilayer would be by increasing the supercell size (which would be analogous to decrease the defect concentration in real systems). This increases the defect-defect distance, so the bandwidth of the defect band in these cases could be reduced, favoring spin polarization. However, we have not tested this possibility.}. Since the PBE functional we used  in this work does not include Van Der Waals interactions, the relaxed value for the interlayer distance may be different from the one reported here and this effect may be important. On the other hand, test calculations in other configurations (including the case of vacancies, which we will discuss later) indicate that small variations in the interlayer distance do not induce any important modifications in the magnetizations or band structures, so the Van Der Waals interactions play no major role in these cases.

The magnetizations observed in Table \ref{tab:H} come from the full occupation of the spin up defect level in the A1 and B1 cases, while no magnetization is observed in the A2 and B2 cases due to lack of spin polarization (for the relaxed interlayer distance). The formation of the C-H bond during the adsorption effectively removes one electron from the honeycomb lattice, leaving an unpaired $p_z$ electron in it. This electron then occupies one of the spin-split defect levels and induces the magnetization. For this reason, we expect that the corresponding spin density should be localized around the H atom. We will see in the next subsection that this is indeed the case. On the other hand, in the A2 and B2 cases, both spin up and down defect levels are equally, but not fully occupied. The resulting effect then is a shift of the Fermi energy, with a small occupation of conduction band states.

Now we move our discussion to the ABC trilayer. In this case, the pure band structure near the Fermi energy consists of a pair of bands with cubic dispersion, followed by two pairs of higher energy quadratic bands, which are degenerate at the $\Gamma$ point. The presence of inversion symmetry in this trilayer leads to a touching point between the valence and conduction bands, which lies outside of the $\Gamma$ point due to trigonal distortion. The hydrogen adsorption lifts this degeneracy and a small gap ($< 0.1$ eV) is opened between the cubic bands. In all cases, the defect levels exhibit spin polarization and, in the A1 and B1 cases, a small spin polarization is also seen in the cubic bands in the vicinity of the $\Gamma$ point. This effect is related to the high density of states associated with a cubic dispersion, due to the almost flat portion near this point. When the Fermi energy lies close to the flat portion, as is the case for the A1 and B1 sites, spin polarization occurs. Similar to the ABA trilayer, a large spin polarization is observed for the defect level in the B1 case, which leads to a full magnetization. In the other cases, the spin up and down levels almost touch at the $\Gamma$ point and a small occupation is observed for the spin down level, thus reducing the magnetization. In the A2 and B2 cases, there is also a shift in the Fermi level and a small occupation of conduction band states, again in similarity with the ABA trilayer.

\begin{figure}[H]
\centering
\includegraphics[width=8.0cm]{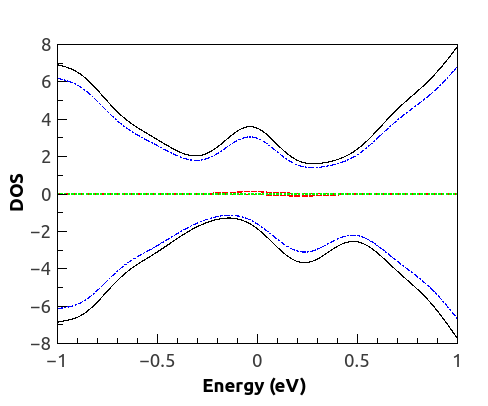}
\caption{\label{fig:h:pdos} Projected Density of States (PDOS) for hydrogen adsorption on top of an A1 site in ABA trilayer graphene. Other configurations show similar features. Black, red dashed, green dotted and blue dash-dotted lines represent the total DOS, contributions from the $s$ orbital of hydrogen, $\sigma$ and $p_z$ orbitals of carbon, respectively. Positive (negative) values represent the spin up (down) PDOS. The Fermi energy is set to zero and a gaussian broadening of $0.2$ eV was used on the energy levels. Contributions from polarized orbitals are not shown.}
\end{figure}

Finally, to finish this subsection, we now discuss the orbital nature of the defect level. To this end, we have performed projected density of states calculations (PDOS) in the vicinity of the Fermi energy. In Fig. \ref{fig:h:pdos}, we show the PDOS for one particular configuration: adsorption on top of an A1 site in the ABA trilayer. All configurations show similar features, including those of the ABC trilayer, so we concentrate our discussion in this case. As can be seen on the figure, the main contribution to the total DOS (black line) near the Fermi energy comes from the $p_z$ orbitals of carbon (blue dash-dotted line), as expected. The central peaks are indications of the defect levels, since they are mostly flat and thus strongly contribute to the DOS. Moreover, the spin up and down peaks are split in energy, indicating spin polarization. Therefore, the defect level features observed in Fig. \ref{fig:h:bands} are correctly reproduced by the PDOS, which also shows that this level has mainly a $p_z$ character, with a small contribution from the $s$ orbital of the H atom.

\subsection{Spin density and local density of states \label{sec:h:spin-ldos}}

\begin{figure}[H]
\centering
\includegraphics[width=7.0cm]{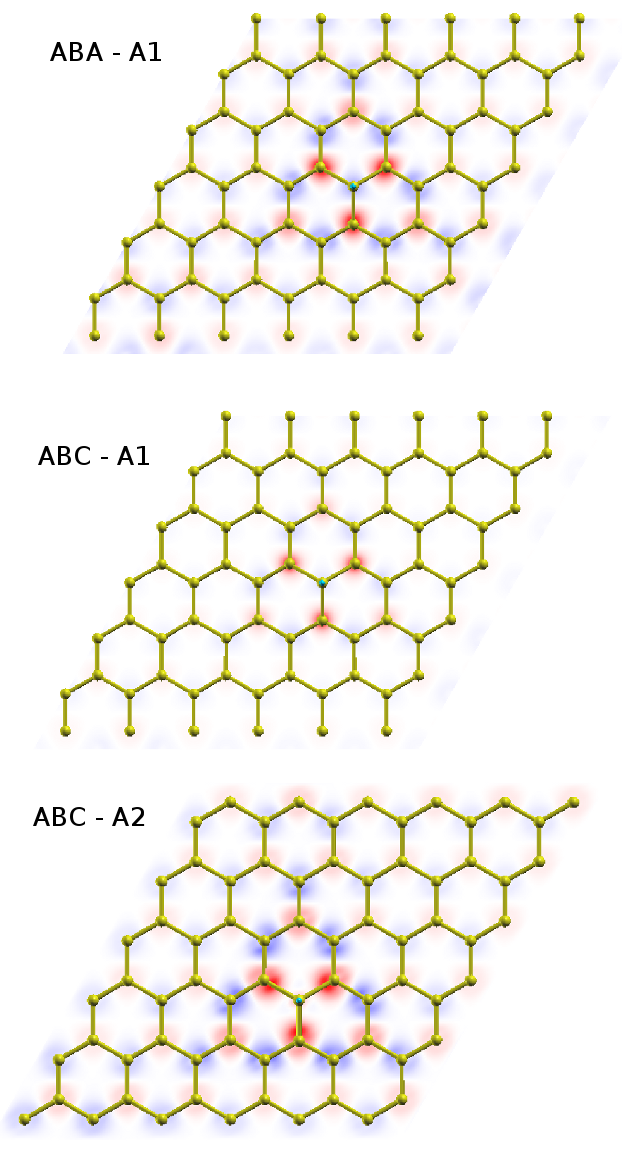}
\caption{\label{fig:h:spin} Spin density for selected configurations. Red (blue) regions indicate spin up (down) excess. The A1 and A2 cases share similar features with the B1 and B2 cases in the corresponding trilayers. For clarity, only the layer containing the adsorbed H atom is shown in each case.}
\end{figure}

In Fig. \ref{fig:h:spin} we show the spin density profiles for selected defect configurations, where only the layer containing the defect is shown. The A2 and B2 cases in the ABA trilayer are not shown in the figure, since they are found to be non-magnetic, as already discussed in the last section. As we can see, in all cases the spin density has a triangular pattern with opposite values for each sublattice, and it is concentrated near the defect site, in agreement with previous calculations on monolayer graphene \cite{oleg_prb, oleg_prl}. The strongest magnitude is observed near the carbon atoms which are first neighbors of the defect site, while the magnitude in the defect site itself is small, with the lowest value observed in the ABC-A2 case. The ABC-A1 case shows a particularly concentrated spin density, while in the other cases the density spreads further into the supercell.

\begin{figure}[H]
\centering
\includegraphics[width=9.0cm]{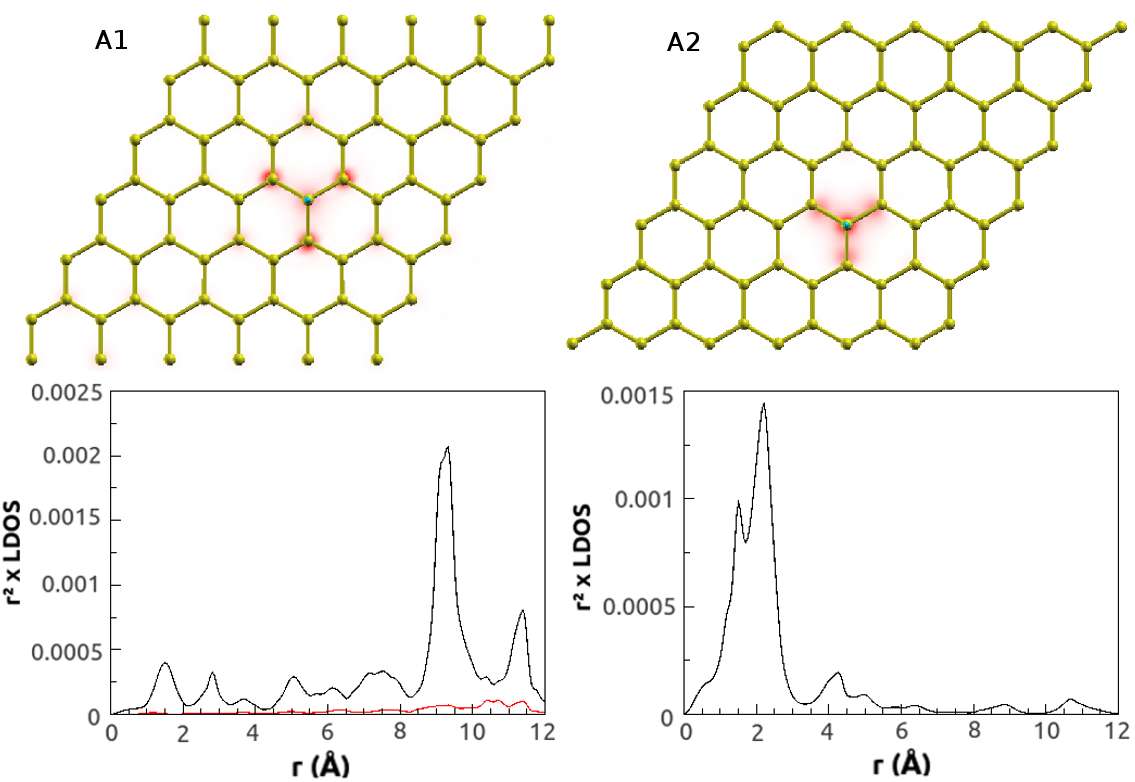}
\includegraphics[width=9.0cm]{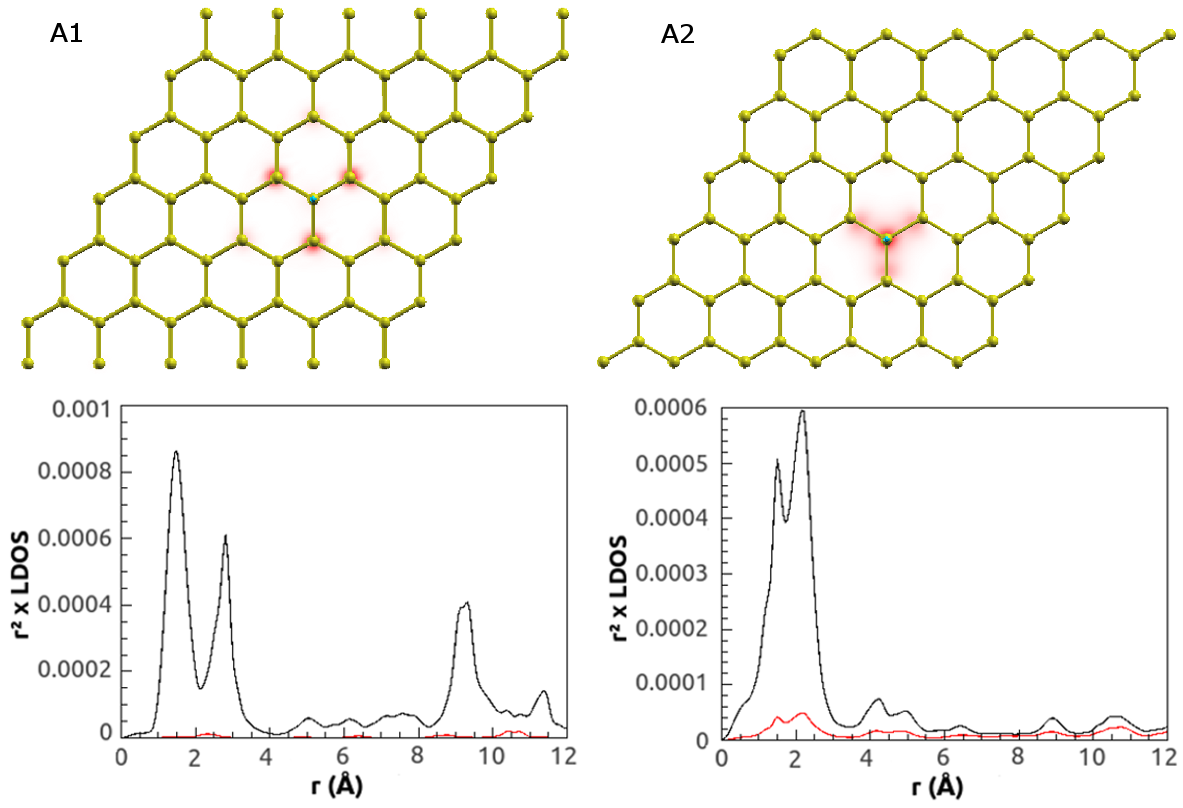}
\caption{\label{fig:h:ldos} Localized density of states for the A1 and A2 configurations in the ABA (top) and ABC (bottom) trilayers, integrated along the bandwidth of the corresponding spin up defect level. For each case, the top panels show the profiles in the layer containing the defects and the bottom panels show the corresponding plots of $r^2 \times$ LDOS as a function of $r$, the distance from the defect center, where the LDOS is averaged over the polar coordinate.}
\end{figure}

Fig. \ref{fig:h:ldos} shows the local density of states (LDOS) for the A1 and A2 configurations in the ABA (top) and ABC (bottom) trilayers. As in the spin density, the results for the B1 and B2 sites are very similar to those of the A1 and A2 sites, so we do not show them here. For each case, the LDOS was integrated in energy between the minimum and maximum values of the spin up defect levels shown in Fig. \ref{fig:h:bands}, so it shows essentially the LDOS associated to the localized defect state. In the graphics panels, we plot $r^2 \times$ LDOS versus $r$, where $r$ is the distance to the defect center and the LDOS is averaged over the polar coordinate. For a quasi-localized state, the associated wavefunction decays as $1/r$, as predicted by tight-binding calculations \cite{pereira_prl, nanda_njp} and observed in STM experiments for vacancies in graphite \cite{ugeda_prl}. Therefore, the magnitude of the LDOS peaks in a quasi-localized state would decay as $1/r^2$ and $r^2 \times$ LDOS should be a constant at these peaks. We look for this possibility in our calculations. For both stackings, we can see then that, when the defect site lies on the middle layer, the LDOS is more concentrated near the defect center and the peaks do not follow a quasi-localized behavior. On the other hand, when it is on an external layer, the LDOS spreads more throughout the sheet and, particularly, the ABA-A1 case shows three peaks at $r < 6$ \AA \ of roughly the same size, which are compatible with a quasi-localized behavior. In the ABC-A1 case, however, we see from the profile that the magnitude of the LDOS at the defect site is reduced when compared to the previous case (this can also be seen in the spin density profiles of Fig. \ref{fig:h:spin}) and the graph shows that the first two peaks have slightly different magnitudes, so they are not fully compatible with a quasi-localized behavior. In all cases, for $r > 6$ \AA \ (roughly half the supercell dimension), the observed peaks could contain contributions from periodic images of the defect, so we do not consider them in our discussion. We also didn't see any LDOS peaks associated with second neighbors, which agrees with the prediction that the wavefunction amplitude in sites belonging to the same sublattice as the defect site in the monolayer should be small (it would be zero if electron-hole symmetry was present).

\subsection{Dependence with the external field for the ABC trilayer \label{sec:h:field}}

\begin{figure*}
\centering
\includegraphics[width=\textwidth]{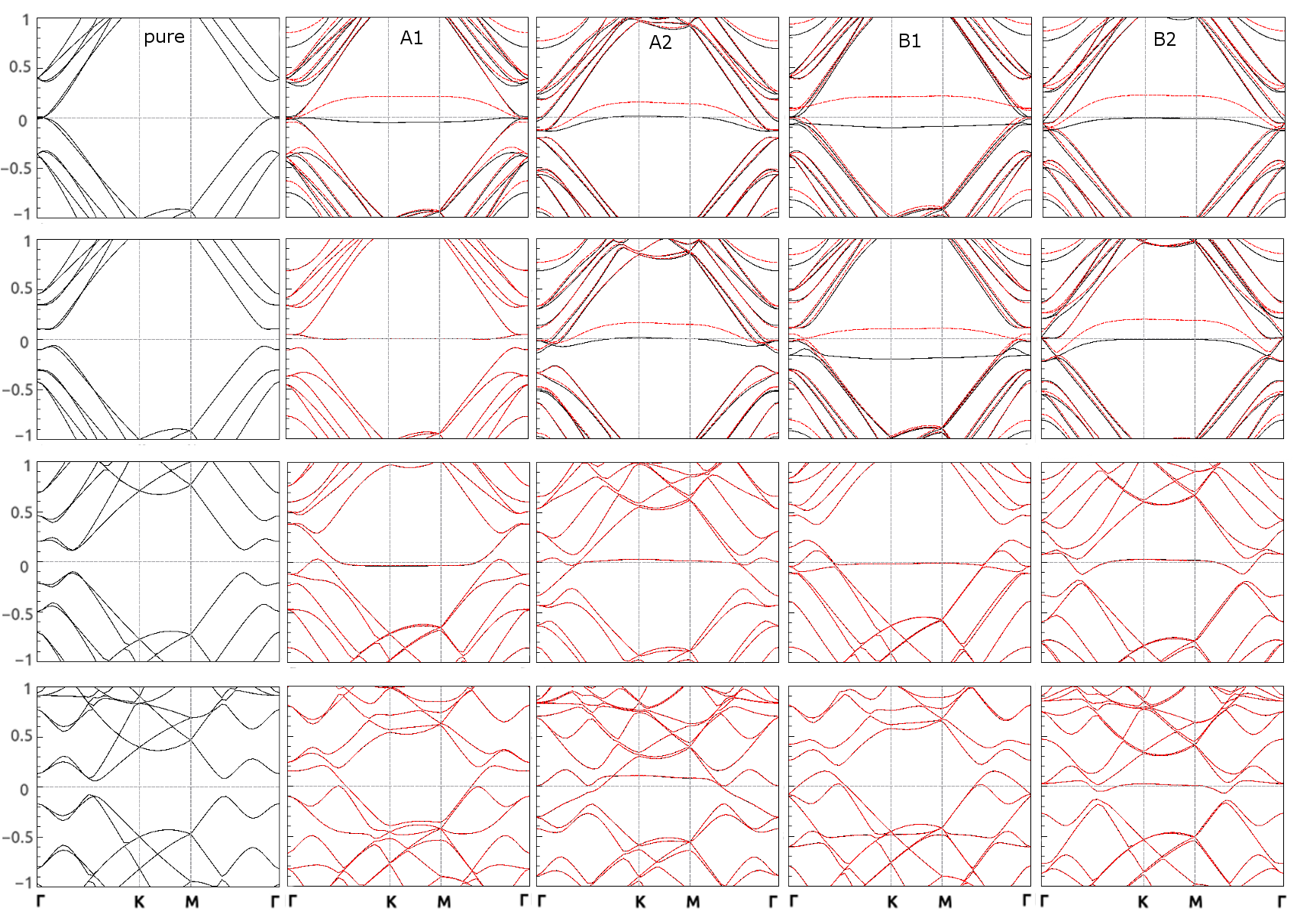}
\caption{\label{fig:h:field} Evolution of the band structure with electrical field for different H adsorption configurations on the ABC trilayer. From top to bottom, the rows correspond to field values $E = 0$, $0.1$, $0.5$ and $1.0$ V/\AA. The columns correspond to the defect configurations indicated in the first row. Black (red dashed) lines represent spin up (down) bands and the Fermi energy is set to zero in all cases.}
\end{figure*}

\begin{figure}[H]
\centering
\includegraphics[width=7.0cm]{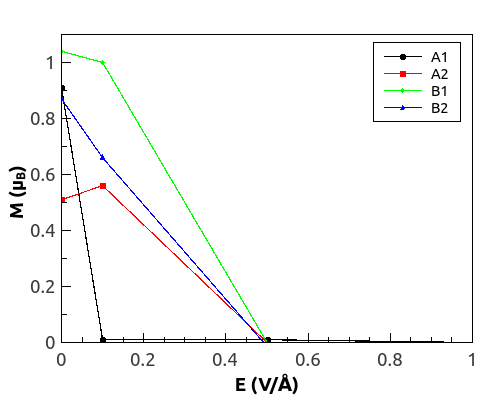}
\caption{\label{fig:h:mag} Evolution of the supercell magnetization with electrical field. Black, red, green and blue lines represent the A1, A2, B1 and B2 defect configurations, respectively.}
\end{figure}

To finish our discussion on hydrogen adsorption, we now discuss the effect of the application of an external electrical field perpendicular to the layers in the ABC trilayer. As we have mentioned before, such field opens a tunable gap in the pure trilayer. In the first column of Fig. \ref{fig:h:field}, we show the evolution of this gap in the pristine supercell for four field values, from top to bottom: $E = 0$ (same as in Fig. \ref{fig:h:bands}), $0.1$, $0.5$ and $1.0$ V/\AA. In the other columns, we show the corresponding band structures for different defect configurations. For small field values such as in the second row, we can see that the bands coming from the pure trilayer evolve in a way similar as they do in the pure supercell, with the opening or increase of small gaps. The only exception is a linear-like band in the B2 case, which crosses the gap between the cubic bands and it is already present in zero field. The defect bands remain close to the Fermi energy and they are slightly changed near the $\Gamma$ point, thus indicating a small increase in the hybridization with band levels. These changes combined with small shifts in their positions lead to small magnetization changes for the A2, B1 and B2 cases, while the A1 case becomes non-magnetic, as shown in Fig. \ref{fig:h:mag}. 

For higher field values, the hybridization between the defect level and the bands becomes much stronger and the connectivity of the bands is modified, also making all configurations become non-magnetic. For $E = 1.0$ V/\AA, we can see remains of the flat dispersion of the defect level mixed below the Fermi energy for the A1 and B1 cases and slightly above it for the A2 case. The change in connectivity also leads to the formation of a nearly flat midgap state in the B2 case, which is degenerate at the $\Gamma$ point. This could indicate the existence of a truly localized state inside the gap, in the sense that the associated wavefunction becomes exponentially localized, as revealed by previous tight-binding calculations \cite{castro_sscomm} \footnote{This previous work actually studies a single vacancy, but the tight-binding model used considers that the only effect of the vacancy is the removal of a $p_z$ electron from the lattice, without considering the removal of the $\sigma$ electrons and the resulting structural distortion (see section \ref{sec:vacancy}). As such, this model is actually more adequate to study Hydrogen adsorption, as supported by our DFT calculations (the bandwidth of the defect bands is smaller).}. However, we don't see any qualitative changes in the LDOS, with the pattern remaining equal to the one observed in Fig. \ref{fig:h:ldos} (bottom right). We also point out that the A1 and A2 cases also show midgap-like states, but they have large bandwidths and are very different from the original defect levels, again due to rehybridization. Finally, we remark again that the difference between the A2 and B2 cases, sites that are originally equivalent by symmetry in the pure trilayer, lies in our choice of placement of the hydrogen atom. In the former, it lies between two carbon atoms in adjacent sheets (the B1 and A2 sites), while in the latter it lies between the B2 site and a hexagon center in the top sheet, thus leading to different calculated band structures.

\section{Isolated Vacancy \label{sec:vacancy}}

\subsection{Structural properties, formation energies and magnetic moments \label{sec:vac:struct}}

\begin{figure}[H]
\centering
\includegraphics[width=8.0cm]{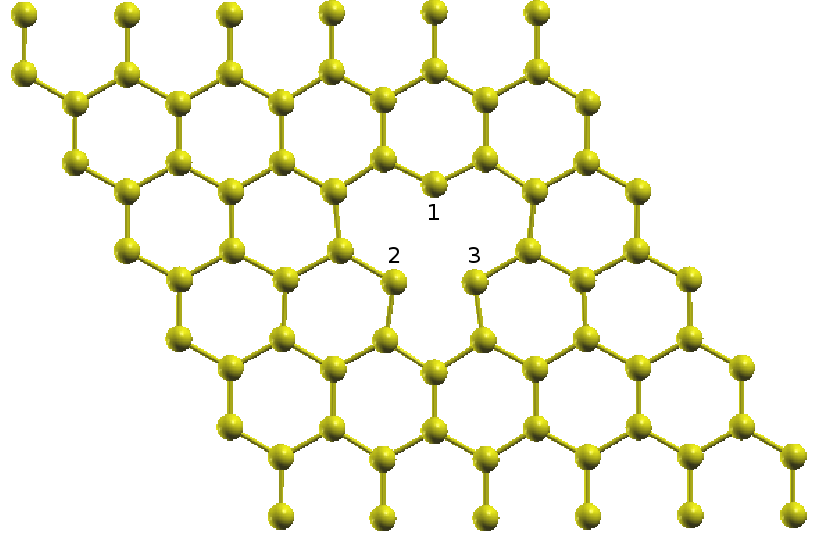}
\caption{\label{fig:vac:relax} Layer containing the vacancy after structural relaxation. The atoms $2$ and $3$ move closer to each other and form a weak $\sigma$ bond, while the atom $1$ remains with an unpaired $\sigma$ electron, which also contributes to the magnetization.}
\end{figure}

Now we move our discussion to the case of an isolated vacancy. Since many of the results we show here are similar to the case of hydrogen adsorption, we will focus our discussion on the differences between the two cases. We begin with the structural properties, which already show important differences. In Fig. \ref{fig:vac:relax}, we show the structure of a layer containing the vacancy after the relaxation. The results are very similar for all defect configurations and both stackings. The removal of a carbon atom from the lattice leaves three unpaired $\sigma$ electrons, one in each neighboring atom (labeled $1$, $2$ and $3$ in the figure), along with an unpaired $p_z$ electron already seen in H adsorption. This leads to a larger structural deformation, where the atoms $2$ and $3$ move closer to each other and form a weak bond, thus saturating their unpaired electrons. The atom $1$, however, remains with an unpaired $\sigma$ electron which, together with the $p_z$ electron, contributes to the magnetization. The local environment of the defect becomes an isosceles triangle, thus breaking the triangular symmetry of the lattice. Such a distortion is known as a Jahn-Teller distortion, and it has already been identified in single layer graphene. The bond lengths for spin-polarized calculations are reported in Table \ref{tab:vac} and are in good agreement with previous calculations on the monolayer, as well as the other properties reported \cite{yuchen_njp,oleg_prb, nanda_njp, faccio_jchem, wang_prb_2012, casartelli_prb}. Here we have not considered the B2 case in the ABC trilayer, since the A2 and B2 sites are equivalent by symmetry in the pristine trilayer and as such we do not expect any noticeable differences between them. For unpolarized calculations only, we have also observed an out-of-plane distortion in atom $1$, which moves about $0.5$ \AA \ in that direction, with the exception of the A2 and B2 cases in the ABA trilayer, where the distortion remains purely planar. Such a difference between spin polarized and unpolarized calculations has also been reported previously \cite{yuchen_njp, nanda_njp}. 

Of course, there are three equivalent distorted configurations, since the atoms $1$ and $2$, or $1$ and $3$ could also form the bond, leaving the remaining atom with the unpaired electron. This could lead to a dynamical Jahn-Teller effect, where the system could be constantly changing between these configurations, if the energy barrier between them is low enough compared with the temperature \cite{nanda_prb}.

\begin{table}[H]
\caption{\label{tab:vac} Results for spin-polarized calculations of an isolated vacancy on trilayer graphene: bond lengths of the first neighbors of the defect site (atoms $1$, $2$ and $3$ in Fig. \ref{fig:vac:relax}), formation energies ($E_f$) and magnetic moments per unit supercell ($M$) for each configuration studied. For all cases, $d_{12} = d_{13}$.}
\begin{tabular}{  *{5}{C{1.6cm}} }
\hline
Defect Site & $d_{23}$ (\AA) & $d_{12}$ (\AA) & $E_f$ (eV) & $M$ ($\mu_B$) \\
\hline
\multicolumn{5}{c}{ABA trilayer} \\
\hline
A1 & 1.95 & 2.56 & 7.80 & 1.46 \\
A2 & 1.91 & 2.54 & 7.81 & 1.34 \\
B1 & 1.93 & 2.55 & 7.83 & 1.43 \\
B2 & 1.94 & 2.55 & 7.77 & 1.43 \\
\hline
\multicolumn{5}{c}{ABC trilayer} \\
\hline
A1 & 1.98 & 2.55 & 7.81 & 1.56 \\
A2 & 1.90 & 2.54 & 7.80 & 1.41 \\
B1 & 1.92 & 2.54 & 7.84 & 1.42 \\
\hline
\end{tabular}
\end{table}

In Table \ref{tab:vac}, we have also included the formation energies and supercell magnetizations for each vacancy configuration. In this case, the formation energies are defined as
\begin{equation}\label{eq:form_en_vac}
E_f = E_{trilayer + vac} - \frac{N-1}{N} E_{trilayer},
\end{equation}
where $E_{trilayer + vac}$ is the total energy of the supercell containing the defect, $E_{trilayer}$ is the total energy of the pristine supercell and $N$ is the number of atoms in that supercell (in our case, $N = 72$ in a $6 \times 6$ supercell). Comparing Tables \ref{tab:H} and \ref{tab:vac}, we see that the vacancies have a much smaller sensitivity to the site where they are placed, with values that differ by a few tenths of meV. This could be related to the large structural distortion they induce, which modifies the local environment in the layer and is also insensitive to the defect site. However, for unpolarized calculations, the formation energies are $0.25 - 0.5$ eV higher and show a larger layer sensitivity, which is related to the presence (or lack) of out-of-plane distortion in atom $1$, as discussed above.

On the other hand, the magnetizations show a large sensitivity to the defect site. Sites that have one C atom right on top of them (B1 in both trilayers and also the A2 site in the ABC trilayer) have roughly the same magnetization, while the A1 sites, which have hexagon centers on top of them have larger magnetizations. The B2 site in the ABA trilayer, which has one hexagon center above and one below it has a magnetization similar to those of the first group, while the A2 site of the same trilayer, which has one site above and one below it, has the smallest magnetization. This indicates that the mechanism responsible for the observed trends could be a charge transfer between the layers, which changes the occupation of the defect levels and reduces the magnetization. The interlayer coupling is stronger in sites that are on top of each other, so the charge transfer could be more intense in these cases (and specially intense in the ABA-A2 case). We now discuss these results in terms of the band structure.

\subsection{Band structure \label{sec:vac:bands}}

\begin{figure*}
\centering
\includegraphics[width=\textwidth]{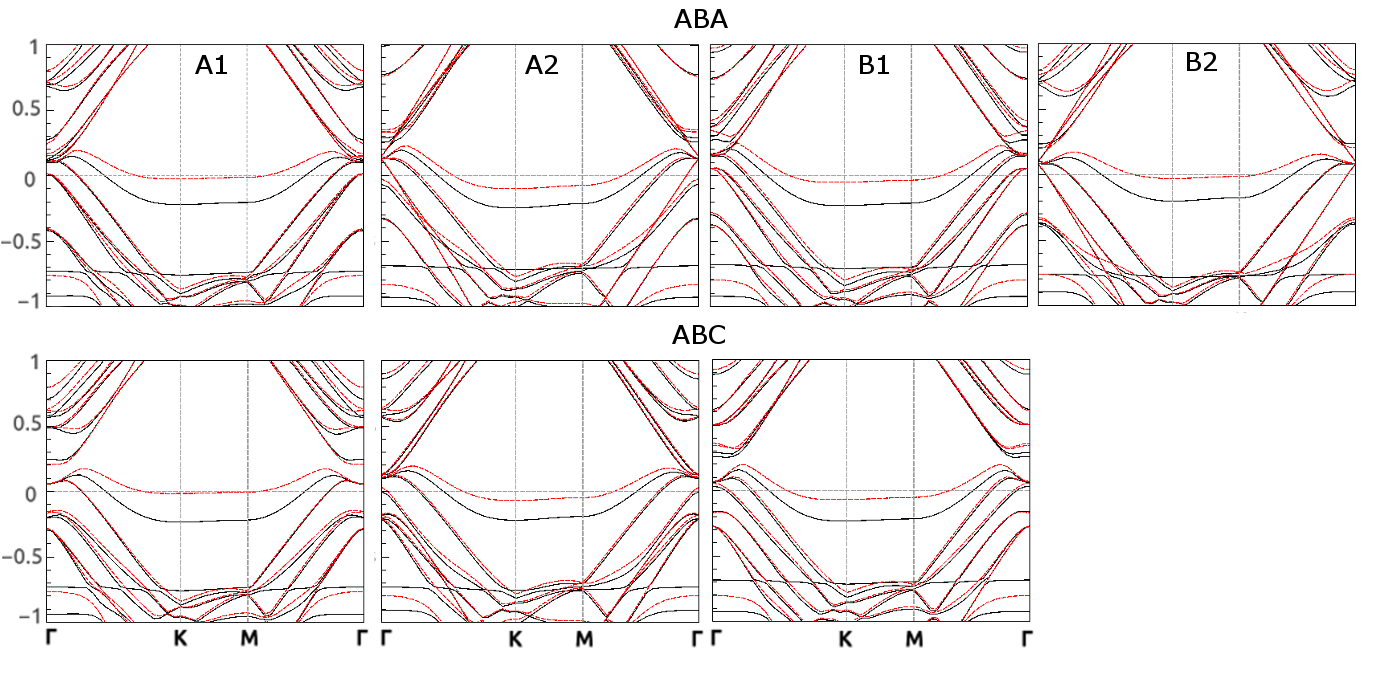}
\caption{\label{fig:vac:bands} Band structure results for an isolated vacancy on ABA (top) and ABC (bottom) trilayer graphene. The vacancy is on the site indicated in each frame. Spin up (down) levels are represented by black (red dashed) lines and the Fermi level is set to zero in all cases. Energies are in eV. For reference, the band structures of the pristine trilayers are shown in Fig. \ref{fig:h:bands} (first column).}
\end{figure*}

In Fig. \ref{fig:vac:bands}, we show the band structures for each vacancy configuration. As in the hydrogen adsorption case, we can see the presence of defect levels near the Fermi energy, which are mainly of $p_z$ character, as we shall see below. We can also see the presence of a flat spin up level below the Fermi energy, between $-0.8$ and $-0.7$ eV, which was absent in the adsorption case. This level has $\sigma$ character, and comes from the unpaired $\sigma$ electron. It suffers a large exchange split, with the corresponding spin down level lying more than $1.0$ eV above the Fermi level. Therefore, this spin up $\sigma$ level contributes a full $1 \mu_B$ to the magnetization. The other two unpaired $\sigma$ electrons also yield defect levels but, since they form a weak bond, these levels lie far apart from the Fermi level, the exchange splittings are small and hence they do not contribute to the magnetization. The remaining contribution to the magnetization comes from the $p_z$ levels, which have larger bandwidths than those observed in the adsorption case. This indicates a larger interaction with band states, leading to charge transfer between band and defect levels, partial occupation of both spin up and down levels and a reduced contribution to the magnetization ($< 1.0 \mu_B$) in all cases. In particular, the different occupations of the spin down $p_z$ level in each case are the main responsible for the trends observed in Table \ref{tab:vac}. This level is occupied in expense of both spin up $p_z$ level and valence band states, leading to a Fermi level shift (with respect to the value in the pure trilayers), which in turn could indicate a charge transfer between the layers, as we have mentioned before.

We now discuss the behavior of the $p_z$ levels in each case in more detail. In the ABA trilayer, we can see that when the vacancy lies on the middle layer (A2 or B2 sites), the linear bands from the pristine trilayer are preserved, while a sizable gap of $0.1 - 0.2$ eV is observed for the quadratic bands. However, when the vacancy is on a external layer, the inverse situation happens, with the quadratic bands preserved and a large gap in the linear bands. This could be related to the mirror symmetry of the C $p_z$ orbitals in each case. Tight-binding calculations in the pristine trilayer show that A2 and B2 orbitals only contribute to the quadratic bands, while A1 and B1 orbitals contribute both to linear and quadratic bands, by combining with A3 and B3 orbitals with odd or even combinations with respect to mirror symmetry, respectively \cite{McCannABA}. Therefore, a vacancy in the A2 or B2 sites should not have an important effect on the linear bands, while a vacancy in the A1 or B1 sites may not affect the shape of the quadratic bands near the Fermi level. This result is in contrast with we have seen on the H adsorption. There, a defect on the middle layer also leads to a widening of the gap between the linear bands (\ref{fig:h:bands}), which we believe is related to the small $sp^2 - sp^3$ rehybridization induced by the formation of the C-H bond, thus breaking the mirror symmetry in the vicinity of the defect.

In the ABC trilayer, a similar situation happens. The presence of a vacancy in the middle layer does not open a large gap between the cubic bands, while a $0.1$ eV gap can be seen in the A1 and B1 cases. Again, this is related to the symmetry of this trilayer. In this case, tight-binding calculations show that the cubic bands are formed by combinations of $p_z$ orbitals coming from the A1 and B3 sites, where the interlayer coupling is weaker and, as such, they are called low-energy sites \cite{MacDonaldABC}. Therefore, the presence of a defect in the middle layer should not affect the low-energy bands.

\begin{figure}[H]
\centering
\includegraphics[width=7.0cm]{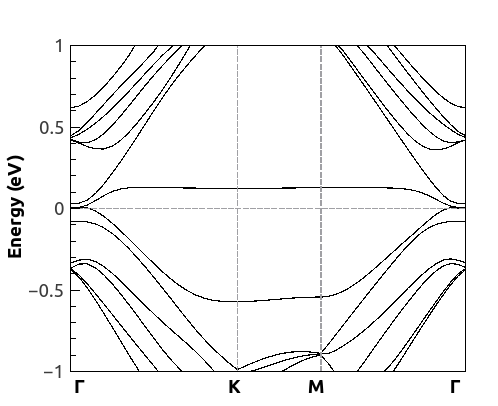}
\caption{\label{fig:vac:bands_nm} Unpolarized band structure for a vacancy on an A1 site in the ABC trilayer. The out-of-plane displacement of atom $1$ in this case leads to rehybridization between the $\sigma$ and $p_z$ levels, resulting in defect levels with increased bandwidths. Other cases show similar features, with the exception of A2 and B2 cases in the ABA trilayer, shown in Fig. \ref{fig:vac:bands_nm2}.}
\end{figure}

\begin{figure}[H]
\centering
\includegraphics[width=8.8cm]{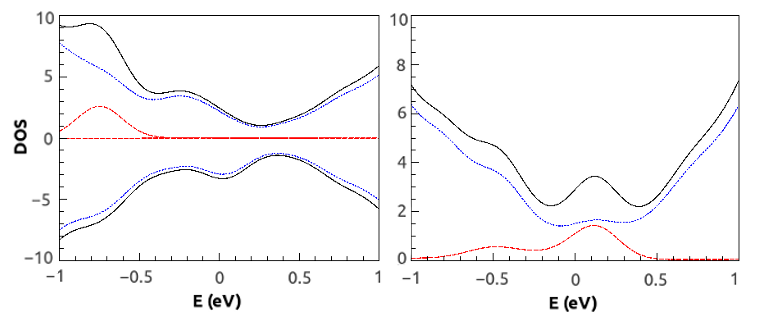}
\caption{\label{fig:vac:pdos} PDOS for a vacancy on an A1 site in ABC trilayer graphene, comparing spin-polarized (left) and unpolarized (right) calculations. Black, red dashed and blue dotted lines represent the total DOS and contributions from $\sigma$ and $p_z$ orbitals of carbon, respectively. Positive (negative) values represent the spin up (down) PDOS. The Fermi energy is set to zero. The peak and shoulder structure on the $\sigma$ curve on the right indicates mixing between the $\sigma$ and $p_z$ defect levels in unpolarized calculations.}
\end{figure}

\begin{figure}[H]
\centering
\includegraphics[width=7.0cm]{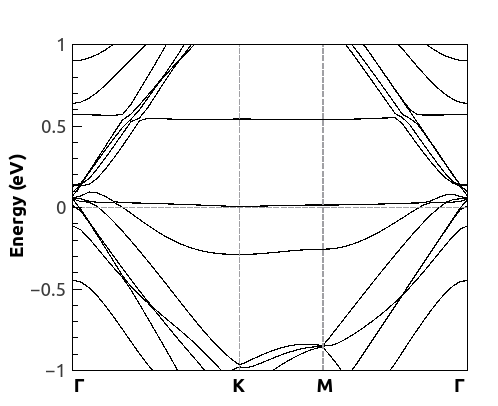}
\caption{\label{fig:vac:bands_nm2} Unpolarized band structure for a vacancy on an A2 site in the ABA trilayer. The $\sigma$ and $p_z$ levels are decoupled in this case due to the lack of out-of-plane displacement of atom $1$, and the corresponding bandwidths are smaller. A second $\sigma$ level, coming from the weak bond between atoms $2$ and $3$, can also be seen within the energy window in this case.}
\end{figure}

For unpolarized calculations, the results are quite different. In Fig. \ref{fig:vac:bands_nm}, we show an example for the A1 site in the ABC trilayer. As we can see, in this case the $\sigma$ defect level coming from the unpaired electron in atom $1$ lies close to the Fermi level. Due to the energy proximity to the $p_z$ level, these levels rehybridize and form the observed structure, with larger bandwidths than those observed in spin-polarized calculations. This analysis is confirmed by the PDOS calculation shown in Fig. \ref{fig:vac:pdos}, where we compare polarized (left) and unpolarized (right) calculations for the same case. In the polarized case, the levels lie far apart and thus retain their pure $\sigma$ and $p_z$ character, while in the unpolarized case the two defect levels near the Fermi energy show contributions from both orbitals, as can be seen by comparing $\sigma$ (red dashed) curves in each case. As we have mentioned before, in most cases of the unpolarized calculations the atom $1$ suffers an out-of-plane displacement which, as in the H adsorption case, promotes a local $sp^2 - sp^3$ rehybridization, thus allowing the mixing between $\sigma$ and $p_z$ levels. When there is no such displacement, as in the A2 case of ABA trilayer (Fig. \ref{fig:vac:bands_nm2}), these levels are decoupled and have smaller bandwidths, compatible with those observed in the polarized calculation.

\subsection{Spin density and local density of states \label{sec:vac:spin-ldos}}

\begin{figure}[H]
\centering
\includegraphics[width=6.5cm]{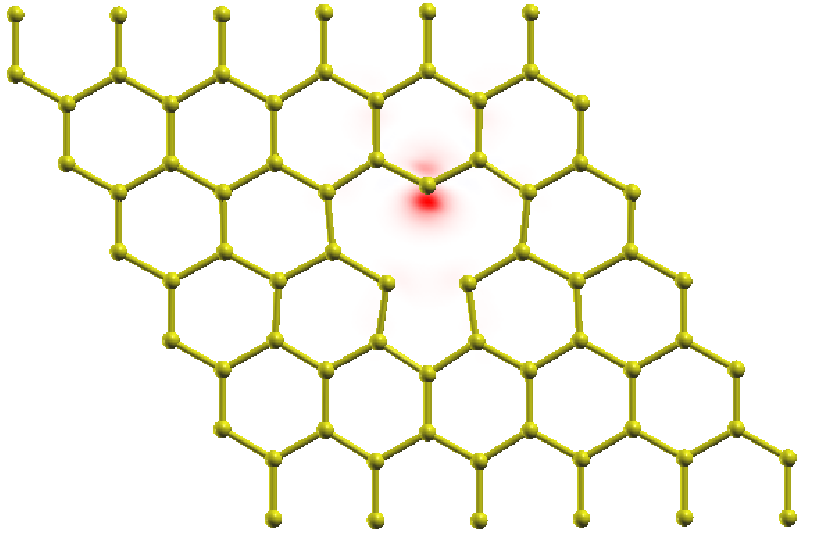}
\caption{\label{fig:vac:spin} Spin density for the A1 case in the ABA trilayer. All other configurations show the same pattern. Red (blue) regions indicate spin up (down) excess. For clarity, only the layer containing the vacancy is shown.}
\end{figure}

We now turn our attention to the induced spin-density. Our calculations show that the patterns are insensitive to the defect site, and they are all equivalent to the pattern shown in Fig. \ref{fig:vac:spin} for the A1 case in the ABA trilayer. This pattern is very different from the triangular pattern observed in the H adsorption case. In this case, the spin-density is very concentrated near atom $1$, which contains an unpaired $\sigma$ electron, as we have mentioned before. This is an indication that the contribution coming from the $\sigma$ defect level to the magnetization is stronger than the one from the $p_z$ levels, which we would expect to induce a triangular pattern similar to the one seen in H adsorption. As we saw in Fig. \ref{fig:vac:bands}, the $\sigma$ level is practically insensitive to the defect site, which explains the observed insensitivity in the spin-density. This pattern is similar to the one observed for vacancies in single layer graphene \cite{oleg_prb, yuchen_njp, faccio_jchem}.

\begin{figure}[H]
\centering
\includegraphics[width=8.8cm]{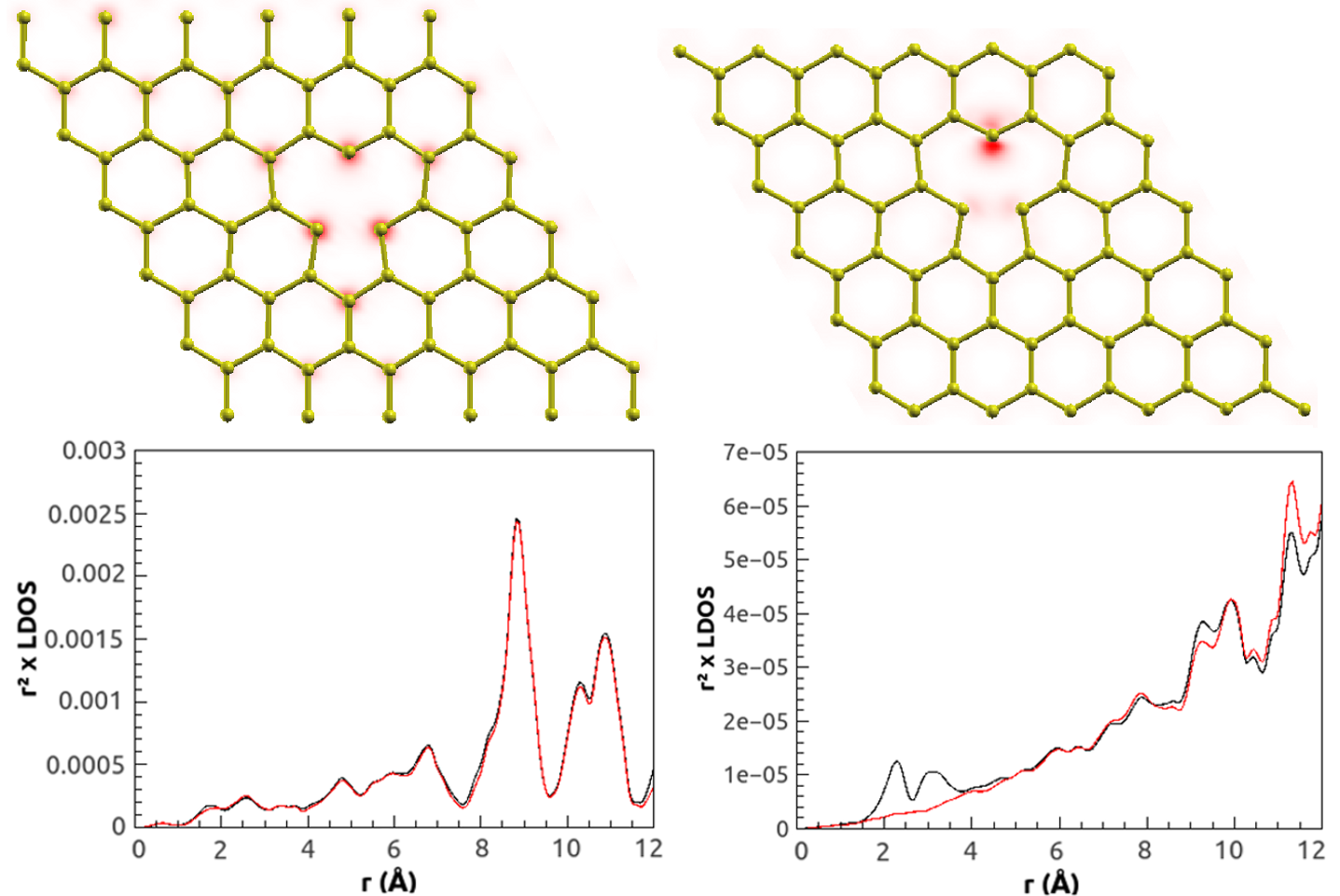}
\caption{\label{fig:vac:ldos} LDOS for the A1 and A2 configurations in the ABA trilayer, integrated along the bandwidth of the $p_z$ defect levels (both spin up and down). For each case, the top panels show the profiles in the layer containing the vacancy and the bottom panels show plots of $r^2 \times$ LDOS as a function of $r$, in the same fashion as Fig. \ref{fig:h:ldos}.}
\end{figure}

On the other hand, the local density of states for the $p_z$ levels in spin-polarized calculations shows a strong layer sensitivity and is roughly stacking independent. In Fig. \ref{fig:vac:ldos}, we compare the results for the A1 and A2 cases in the ABA trilayer, which are similar to the B1 and B2 cases, respectively. The LDOS is integrated along the bandwidth of the $p_z$-up defect level, which also contains the down level within the same energy range (note that the inverse situation occurs in H adsorption, as can be seen in Fig. \ref{fig:h:bands}, so only the spin-up level was included in that case). When the defect is on an external layer, the LDOS shows a triangular pattern, which is more spread out in the supercell than the patterns observed in H adsorption (see Fig. \ref{fig:h:ldos}). The corresponding plot of $r^2 \times$ LDOS also shows more structure and it is roughly constant for $1.5$ \AA $< r < 4$ \AA, which could indicate a quasi-localized behavior in this region. For intermediary distances this behavior is not well satisfied and, for larger distances, larger peaks are observed and they may contain contributions from the periodic images, as was also pointed out in the H adsorption case.

When the defect is in the middle layer, the situation is very different. The LDOS shows a pattern that resembles the spin-density shown in Fig. \ref{fig:vac:spin}. The $r^2 \times$ LDOS plot shows a two-peaked structure around atom $1$, and a small peak around the bond between atoms $2$ and $3$ can also be seen in the profile map. Therefore, the defect level wavefunction is more localized near the vacancy center and does not follow the $1/r^2$ power-law. We have seen a similar situation in the hydrogen adsorption case, where the LDOS is also more localized at the defect center when it is on the middle layer, although the observed patterns are very different. Note also that the LDOS amplitudes of the spin up and down levels (black and red lines in the line profiles) are very different near the peak at atom $1$, but they are very similar for larger distances. This could indicate that this long range contribution, which decays weaker than $1/r^2$ (since $r^2 \times$ LDOS increases with $r$) and has a very weak magnitude when compared to the previous plots, could come from band levels within the integration window, which would add to possible contributions from periodic images. In the other plots, this contribution may be present, but its magnitude is much weaker than the others we have discussed previously.

Finally, we mention that, for unpolarized calculations, we do not see a layer sensitivity and all patterns are similar to the localized pattern we have just discussed. This could be related to the presence of a $\sigma$ level near the Fermi energy in this case which, as we have seen, rehybridizes with the $p_z$ level when the defect is on an external layer. The corresponding wavefunctions then have a contribution from the $\sigma$ orbital coming from the unpaired electron, leading to a stronger localization.

\subsection{Dependence with the external field for the ABC trilayer \label{sec:vac:field}}

\begin{figure*}
\centering
\includegraphics[width=13cm]{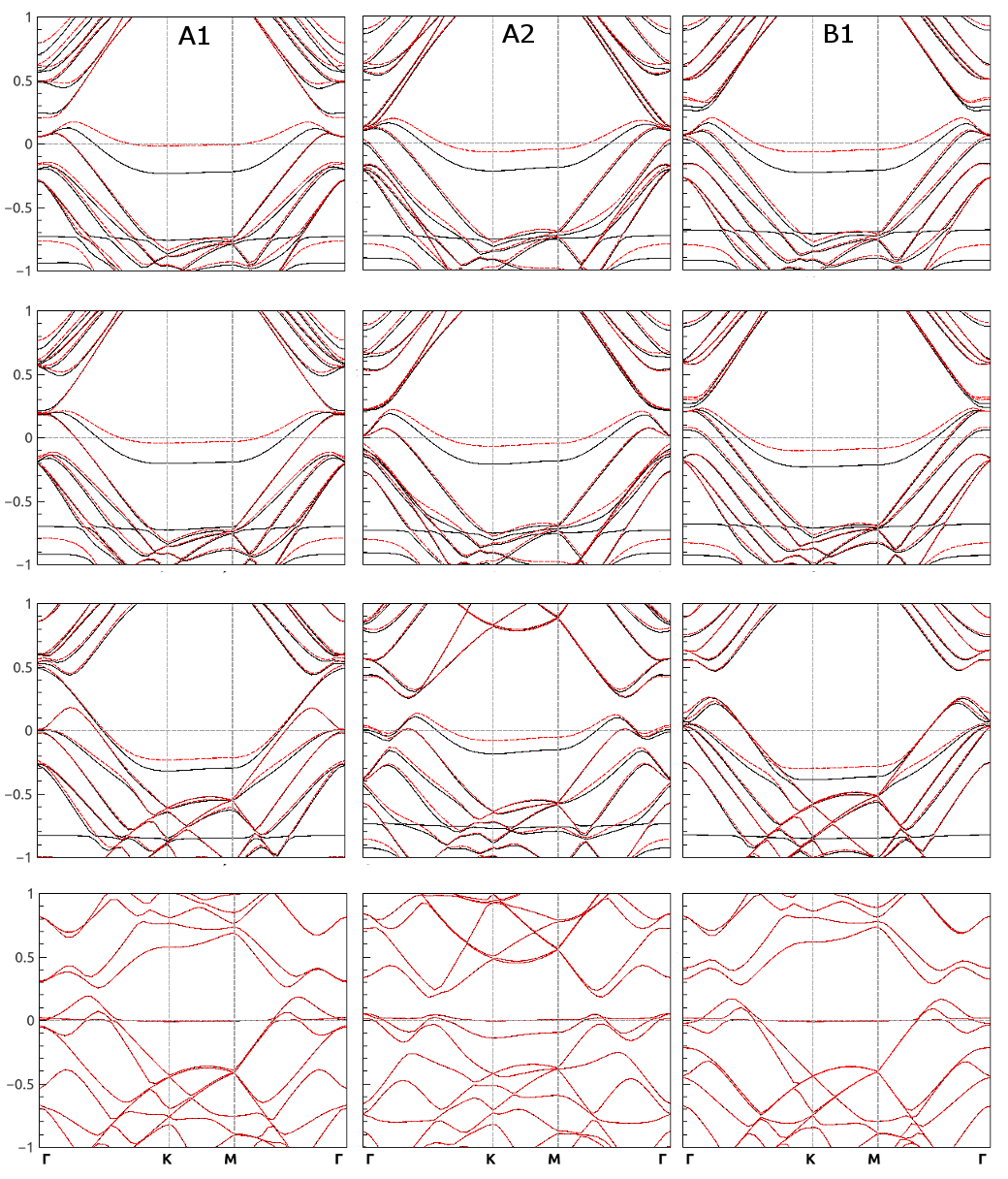}
\caption{\label{fig:vac:field} Evolution of the band structure with electrical field for different vacancy configurations on the ABC trilayer. From top to bottom, the rows correspond to field values $E = 0$, $0.1$, $0.5$ and $1.0$ V/\AA. The columns correspond to the defect configurations indicated in the first row. Black (red dashed) lines represent spin up (down) bands and the Fermi energy is set to zero in all cases. Band structures for the pure trilayer are shown in Fig. \ref{fig:h:field}.}
\end{figure*}

We now discuss the effect of the application of an external electrical field in the ABC trilayer with a vacancy. In Fig. \ref{fig:vac:field}, we show the band structures for zero field and three field values for each vacancy configuration, as we did in Fig. \ref{fig:h:field}. The trends observed here are very similar to those of H adsorption. For a small field value (second row), the band levels remain similar to those of the pure trilayer, but the gap between the cubic bands is affected by the vacancy. For the A1 and B1 configurations, the gap is reduced and almost closes, indicating that the charge transfer between the layers induced by the field is compensated by an opposite charge transfer due to presence of the vacancy (the field is applied in the direction of layer 1 to layer 3). A similar behavior is seen on bilayer graphene, when potassium is adsorbed on top of a layer when the system has residual doping \cite{ohta_science}. For the A2 case, however, the gap is very small at zero field and it evolves naturally as the field increases. The small gap at zero field in this case indicates that the external layers contribute almost equally to the charge transfer to the middle layer in this case, so they have roughly the same electrical potential. When they have the same potential, no gap is induced, in agreement with our earlier discussion in section \ref{sec:vac:bands} in terms of the symmetry of the ABC trilayer and the $p_z$ orbitals.

\begin{figure}[H]
\centering
\includegraphics[width=7.0cm]{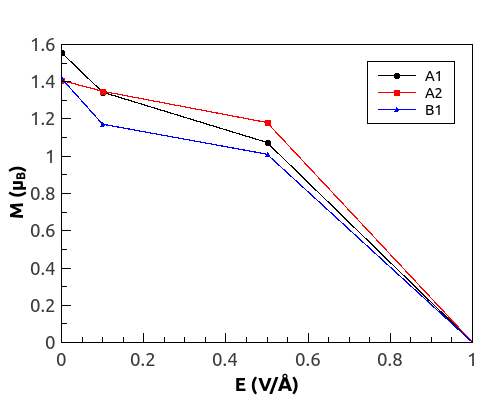}
\caption{\label{fig:vac:mag} Evolution of the supercell magnetization with electrical field. Black, red and blue lines represent the A1, A2, B1 defect configurations, respectively.}
\end{figure}

For larger field values (third and fourth rows), the $p_z$ defect levels strongly mix with the band levels, changing the band connectivity and reducing the magnetization (Fig. \ref{fig:vac:mag}). For $E = 0.5$ V/\AA, we can see a "midgap" state with a large bandwidth in the A1 and B1 cases and a small one in the A2 case, but none of them lie completely inside the gap between the adjacent bands. By increasing even more the field, the spin polarization eventually disappears and the sigma level is pulled back to the Fermi level, as we have discussed in section \ref{sec:vac:bands} and can be seen in the last row. The $p_z$ defect level (now rehybridized with band levels) remains visible only in the A2 case, and still has a small bandwidth. The corresponding LDOS for this level has the same localized pattern shown in Fig. \ref{fig:vac:ldos} (right), so this case could be a candidate for a truly localized midgap state, as predicted by a previous tight-binding calculation \cite{castro_sscomm}. We remark that a similar situation happened in H adsorption for the B2 case, as we have discussed in section \ref{sec:h:field}. We expect this case to actually resemble more the result of a typical tight-binding calculation since, as we have discussed, H adsorption induces small structural changes and remove only a $p_z$ electron from the lattice (tight-binding calculations in graphene and related systems usually neglect structural distortions and include only $p_z$ electrons).

\section{Conclusions \label{sec:conclusions}}

We have studied the electronic and structural properties of hydrogen adsorption and single vacancies on trilayer graphene. We have verified that most properties, such as formation energies, magnetization and local density of states, show a sensitivity to the layer in which the defect is placed, followed by smaller sensitivities to sublattice and stacking type. Both types of defects remove a $p_z$ electron from the lattice, thus inducing a defect level in the band structure near the Fermi energy. This level shows a bandwidth, as a result of the interaction with the band levels near the $\Gamma$ point of the supercell BZ (both $K$ and $K'$ points from the primitive cell BZ are folded into this point), which is larger for the vacancy case. For spin polarized calculations, this level also suffers an exchange splitting due to the zero energy instability (large density of states at the Fermi energy due to the flat portion of the level), and the different populations of the spin up and down levels lead to the different magnetizations observed in each defect configuration. For vacancies, the removal of a carbon atom also leaves three unpaired $\sigma$ electrons, two of which form a weak bond, while the third remains unpaired, resulting in a Jahn-Teller structural deformation. These electrons also induce defect levels, one of which lies close to the Fermi level (the one coming from the unbound electron) and suffers a large exchange splitting, thus also contributing to the magnetization. Unpolarized calculations may show a rehybridization between this level and the $p_z$ level, depending on whether an out-of-plane displacement of the atom containing the unpaired electron occurs.

Spin density calculations exhibit a triangular pattern in H adsorption and a localized pattern in vacancies, where the localization occurs near the atom containing the unpaired $\sigma$ electron, whose contribution is stronger than that of the $p_z$ defect level. Local density of states show patterns that are more localized near the defect center when the defect is placed on the middle layer. Finally, our calculations with an applied external electrical field in the ABC trilayer indicate that the $p_z$ defect level strongly mixes with band levels for high field values, for both type of defects, resulting in reduction and eventually loss of magnetization. There is also a possibility of a midgap-like state with a small bandwidth when the defect is placed on the middle layer, as predicted by previous tight-binding calculations. Such a state can be very important for the transport properties of the ABC trilayer when used in electronic devices.

\begin{acknowledgments}
This work was supported by the Brazilian funding agencies: CNPq, CAPES, FAPERJ and INCT-Nanomateriais de Carbono.
\end{acknowledgments}

\bibliography{Artigo}

\end{document}